\documentclass[sn-nature]{sn-jnl}

\usepackage{graphicx}%
\usepackage{multirow}%
\usepackage{amsmath,amssymb,amsfonts}%
\usepackage{amsthm}%
\usepackage{mathrsfs}%
\usepackage[title]{appendix}%
\usepackage{xcolor}%
\usepackage{textcomp}%
\usepackage{manyfoot}%
\usepackage{booktabs}%
\usepackage{algorithm}%
\usepackage{algorithmicx}%
\usepackage{algpseudocode}%
\usepackage{listings}%
\usepackage{xspace}
\usepackage{dirtytalk}
\usepackage{cleveref}
\newcommand{\bplus}{$B_{1}^{\!+}$\xspace}
\newcommand{\bnull}{$B_0$\xspace}

\MakeRobust{\say}

\theoremstyle{thmstyleone}%
\theoremstyle{thmstyletwo}%

\theoremstyle{thmstylethree}%

\begin{document}
\title{Coaxial Dipole Array with Switching Transmit Sensitivities for ultrahigh field MRI}


\author*[1,2,3]{\fnm{Dario} \sur{Bosch}}\email{dario.bosch@tuebingen.mpg.de}
\equalcont{These authors contributed equally to this work.}

\author[1]{\fnm{Georgiy A.} \sur{Solomakha}}\email{georgiy.solomakha@tuebingen.mpg.de}
\equalcont{These authors contributed equally to this work.}

\author[1]{\fnm{Felix} \sur{Glang}}\email{felix.glang@tuebingen.mpg.de}

\author[4]{\fnm{Martin} \sur{Freudensprung}}\email{martin.freudensprung@uk-erlangen.de}

\author[1]{\fnm{Nikolai I.} \sur{Avdievich}}\email{nikolai.avdievitch@tuebingen.mpg.de}

\author[1,2]{\fnm{Klaus} \sur{Scheffler}}\email{klaus.scheffler@tuebingen.mpg.de}

\affil*[1]{\orgdiv{Magnetic Resonance Center}, \orgname{Max Planck Institute for Biological Cybernetics}, \orgaddress{\street{Max Plank Ring 8}, \city{T\"ubingen}, \postcode{72076}, \country{Germany}}}

\affil[2]{\orgdiv{Biomedical Magnetic Resonance}, \orgname{University of T\"ubingen}, \orgaddress{\street{Otfried-M\"uller-Straße 51}, \city{T\"ubingen}, \postcode{72076}, \country{Germany}}}

\affil[3]{\orgdiv{Core Facility MRT of the Medical Faculty}, \orgname{University of T\"ubingen}, \orgaddress{\street{Otfried-M\"uller-Straße 51}, \city{T\"ubingen}, \postcode{72076}, \country{Germany}}}

\affil[4]{\orgdiv{Institute of Neuroradiology}, \orgname{University Hospital Erlangen}, \orgaddress{\street{Schwabachanlage 6}, \city{Erlangen}, \postcode{91054}, \country{Germany}}}

\abstract{

\textbf{Purpose:}\\
To investigate dipole antennas with electronically switchable transmit field patterns to improve flip angle homogeneity in ultra-high field MRI

\textbf{Methods:}\\
An array of eight coaxial dipoles with electronically switchable \bplus field profiles was constructed. Alteration of the field profiles was accomplished by modulating the currents along the dipoles using a combination of PIN diodes and lumped inductances.
The behavior of these reconfigurable elements was studied in numerical electromagnetic simulations and 9.4T MRI measurements, investigating rapid switching of transmit sensitivities during excitation pulses in both single-channel and pTx mode operation.

\textbf{Results:}\\
For the simulated dipole elements, modulating the current densities along the dipole's axis causes a $\sim$30\% change of the \bplus field between superior and inferior regions of the brain. When rapidly switched during excitation pulses, this degree of freedom can improve flip angle homogeneity, e.g. by a factor of $\sim$2.2 for a two kT points pTx pulse. For the constructed prototype array, the switching effect was observable but weaker, causing $\sim$10\% superior-inferior \bplus variation.

\textbf{Conclusion:}\\
The proposed coaxial dipole array with switchable transmit sensitivities offers a novel degree of freedom for designing excitation pulses. The approach has the potential to improve flip angle homogeneity without necessitating an expensive increase in the number of independent transmit channels.
\\

\textcolor{red}{
}

}

\keywords{RF-shimming, pTx, kT-points, dipoles, arrays}



\maketitle
\section{Introduction}

In ultra-high field MRI (UHF MRI, $\ge$ 7 Tesla), the transmit (Tx) radio-frequency (RF) field (usually referred to as \bplus field) is inherently inhomogeneous due to its short wavelength compared to the sample size \cite{ladd2018pros,cao2015numerical}. To compensate these inhomogeneities, state-of-the-art UHF MRI systems are often equipped with parallel transmit (pTx) systems \cite{deniz2019parallel}. These systems allow driving a set of Tx RF-elements independently with usually 8 or 16 independent channels \cite{ladd2018pros}, thus giving control over the interference pattern of the RF field generated by the Tx elements by adapting the magnitude and phase of the waveform delivered to each individual element. This is generally referred to as "RF shimming" \cite{mao2006exploring}.
Having individual RF amplifiers for each Tx channel allows modifying the RF magnitude and phase dynamically, as opposed to building a dedicated power splitter for an array of Tx elements that could provide only one fixed phase and amplitude excitation for the array. This is employed, for example, in slice-by-slice RF shimming, or for optimized composite pulses such as spokes pulses or kT-points \cite{cloos2012kt}, where an optimized RF shim is computed for each individual sub-pulse.

The number of degrees of freedom in this process can be increased by increasing the number of Tx channels of the MRI system. However, due to the high technical complexity and cost, very few MRI systems are equipped with more than 8 Tx channels. If an RF-coil with 16-channels is available, 8 elements can be driven by a power splitter symmetrically or asymmetrically \cite{yan2018ratio-adjustable,sappo2024optimizing}, so that 2 or more elements are driven by each single Tx channel. This does not increase the number of degrees of freedom, since the power splitters have a fixed behavior defined by their hardware. It does, however, allow for improved Tx array designs that offer, for example, larger coverage.
Another way of driving a large number of Tx elements is Tx multiplexing \cite{avdievich_transceiver-phased_2011}. Here, multiple Tx elements are connected to a single Tx channel by an electronically controlled switch that allows routing the RF power to a specific element at a time and change that setting within a very short time, i.e. less than 1 µs. This method, however, introduces additional complexity since not only a larger number of elements need to be built, but also switches that can handle the high RF power and still switch fast and reliably. 

In the past years, multiple projects used receive (Rx) loop \cite{glang2022accelerated} and dipole \cite{nikulin2023reconfigurable,solomakha2024dynamic} elements with switchable sensitivities to boost parallel imaging performance \cite{pruessmann1999sense} and reduce the geometrical factor, which describes the noise amplification due to parallel imaging reconstruction. Additional electronic components within the Rx element allow changing the spatial distribution of its sensitivity. The sensitivities were switched back and forth during the signal acquisition phase of an MRI pulse sequence, multiplexing the two states of the Rx elements. The two Rx sensitivity states were then treated like different coil images in SENSE \cite{pruessmann1999sense} parallel imaging reconstruction.

Inspired by this previous work, we adapted the approach of RF elements with switchable sensitivities to transmit elements. A coaxial dipole element was previously proposed by van Leeuven et al. \cite{van2022coax} and evaluated as array element for 9.4T head imaging by Solomakha et al. \cite{solomakha2024evaluation}. Such an array element has a current distribution that is symmetrical around its feed point, causing the generated \bplus field to have the same symmetry. To control the current distribution, two lumped inductors can be placed at the end of the dipole, between the shield and the central conductor of the coaxial cable. If one end of the dipole is made capacitive (or open circuit) instead of inductive, the current distribution is skewed towards that end, causing the \bplus field to be skewed accordingly. The goal is then to use these distinct time-varying transmit sensitivities as additional degree of freedom for RF pulse design.

In this work, we investigate such reconfigurable transmit dipoles based on numerical simulations, demonstrating the potential of this approach to improve flip angle homogeneity. For that, the transmit sensitivities are rapidly switched in between the sub-pulses of both CP mode and optimized kT points excitation pulses.
In addition, we perform an experimental proof-of-concept with a self-built prototype 8-element coil array applied for human brain imaging.
Preliminary simulation results of this approach have been presented on previous conferences \cite{bosch_txloom_2023, bosch_txloom_2024}.

\section{Methods}
\subsection{Design and Numerical Simulations of the Element}

\begin{figure*}[tbp]
  \centering
  \includegraphics[width=0.9\textwidth]{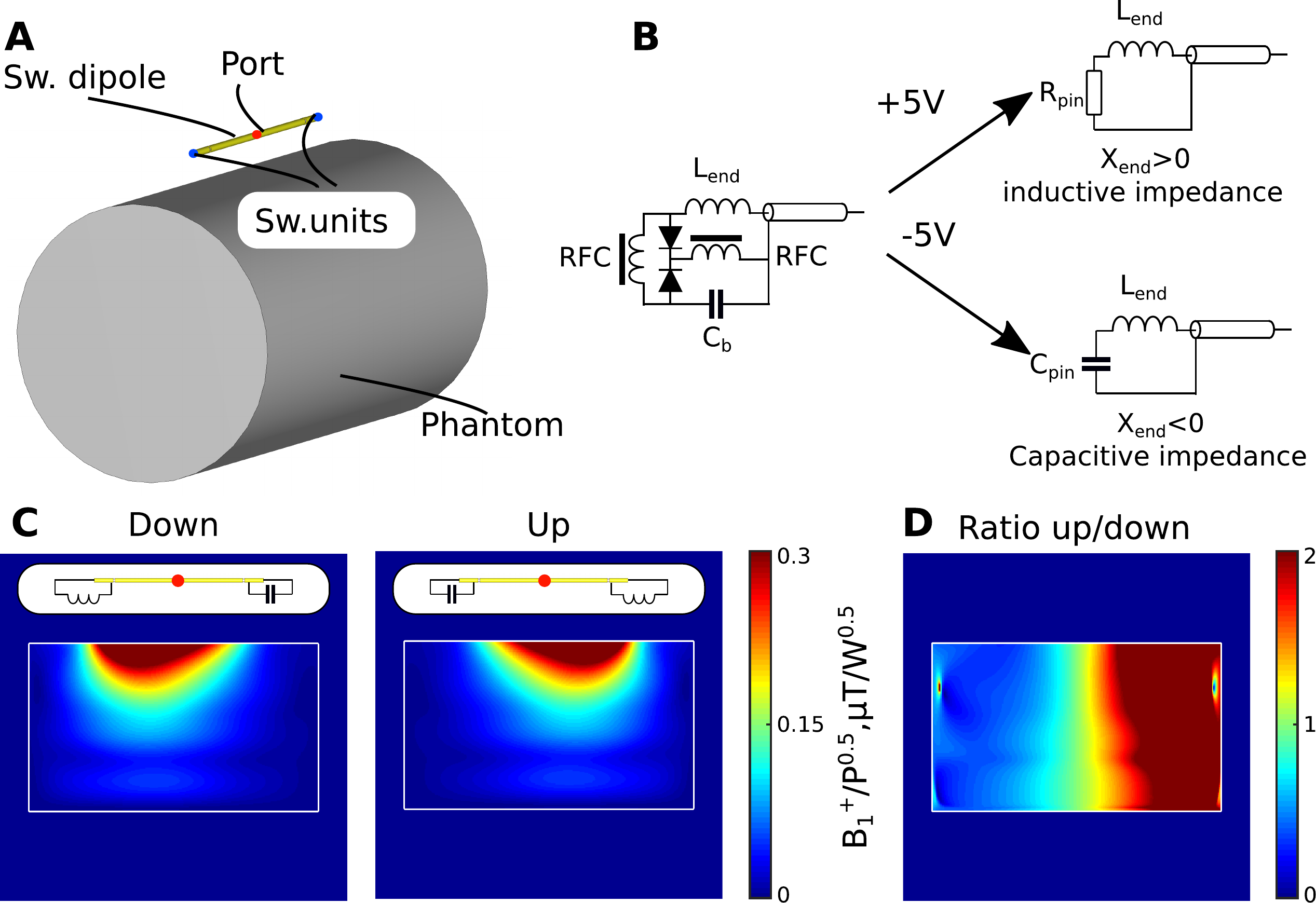}
  \caption{%
    Illustration of functioning principle of the switchable dipole element. \textbf{(A)} View of the numerical model of a single switchable coaxial dipole antenna. \textbf{(B)} Switchable unit circuit for high power operation. \textbf{(C)} Numerically simulated \bplus field for \say{down} and \say{up} configuration of switchable coaxial dipole antenna. \textbf{(D)} Ratio between down and up field configuration.%
    }%
  \label{fig1}
\end{figure*}%

First, single dipoles were simulated to demonstrate how the \bplus field pattern changes with changing load impedance at the dipole's end. This was done in  CST Studio 2021 (Dassault Systèmes, France), using the finite-element method in the frequency domain. A single coaxial dipole was constructed from coaxial cable with a center conductor diameter of 1 mm and a shield diameter of 3.4 mm, similar to previous publications \cite{solomakha2024dynamic,solomakha2024evaluation}. Both central conductor and shield were made from copper. Inner and outer coaxial cable diameters were doubled compared to the real coaxial cable to increase the accuracy of the numerical simulation. The total length of the coaxial dipole was 17 cm with slots located at 20 mm distance from the dipole's ends. The dipole element was loaded to a 175 mm diameter ($D_{ph}$) and 300 mm length ($L_{ph}$) homogeneous cylindrical phantom. The dielectric properties of the phantom ($\epsilon$ = 58, $\sigma$=0.64 S/m) correspond to averaged human brain tissues properties at the 9.4 T MRI frequency of 400 MHz. A view of the numerical model of a single switchable coaxial dipole antenna is presented in \Cref{fig1} (A).

The ideal configuration with \bplus skewed to one side of an object correspond to a dipole antenna with an inductance on that end and an open circuit at the other end. Our previously used switchable unit circuit for reconfigurable receive elements \cite{solomakha2024dynamic} (PIN diode connected in series with inductance) is not suitable for operation in Tx-mode since the negatively biased PIN diode could be opened by the high-voltage RF pulse \cite{Avdievich2007-om}. To prevent the forward biased PIN diode from being switched off, we modified the switching circuit by adding an additional counter-directed PIN diode. The proposed circuit of a high-power switchable unit is presented in \Cref{fig1} (B). In all numerical simulations the following parameters were used for the switching unit circuit: $L_{\text{end}}$=40 nH with Q-factor of 150, $C_{\text{pin}}$=1.1 pF (corresponds to the capacitance of two negatively biased high-power PIN diodes MA4P7446F-1091 (MACOM, USA) in series), $R_{\text{pin}}$= 1 $\Omega$.

\subsection{Coil Design and Simulation}

\begin{figure*}[tbp]
  \centering
  \includegraphics[width=0.95\textwidth]{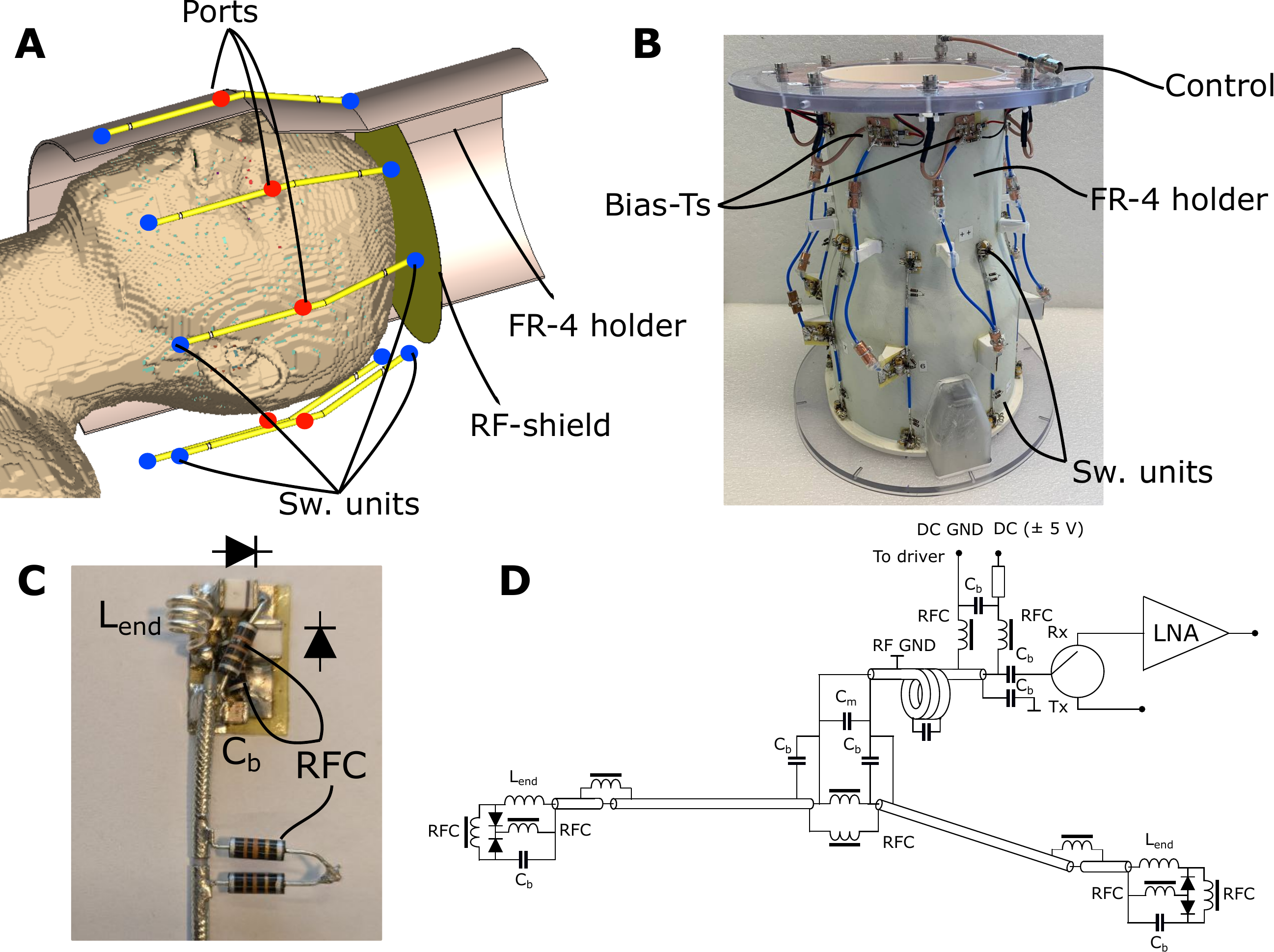}
  \caption{%
     \textbf{(A)} View of the numerical model of the switchable coaxial dipole antenna array in CST Studio 2021 loaded with the Duke voxel model. \textbf{(B)} Photo of the prototype of the switchable coaxial dipole antenna array. \textbf{(C)} Enlarged photo of a switchable unit placed at the coaxial dipole end. \textbf{(D)} Circuit diagram of a switchable coaxial dipole antenna.%
    }%
  \label{fig2}
\end{figure*}

Following simulations of a single switchable dipole, an 8-channel array of switchable coaxial dipole antennas was investigated numerically. For this, the finite integration technique in time-domain was used in CST Studio 2021. The 8-channel switchable array was mounted on a 3\,mm thick FR-4 holder. The shape of the holder was identical to the one previously used for evaluation of a non-switchable coaxial dipole head array for 9.4\,T \cite{solomakha2024evaluation}. A local RF-shield was added at 2\,cm distance from the top end of the dipoles similarly to other dipole arrays \cite{solomakha2024evaluation,avdievich2021folded,nikulin2023double,avdievich2021unshielded} developed in our group to increase field homogeneity. A 1600\,mm length copper cylinder with 640\,mm diameter was added to the numerical model to mimic the RF shield (bore) of our MRI scanner. The array was loaded with multi-tissue human models "Duke" and "Ella" (Zurich MedTech, Switzerland) with 2\,mm resolution. A view of the array's numerical model loaded with the Duke voxel model is presented in Figure \ref{fig2}. 

For comparison, the non-switchable coaxial dipole array from work \cite{solomakha2024evaluation} was also simulated as a reference. The simulations resulted in three sets of \bplus maps: \say{up} and \say{down}, where all sensitivity profiles were shifted in the cranial/caudal direction (\Cref{fig3}), and the reference array's symmetrical maps. Based on the results of the numerical simulations, Q-matrices were calculated and compressed to VOP-matrices \cite{Eichfelder2011-gb} using the modified compression algorithm from Orzada et al. \cite{Orzada2021-fh} with an overestimation factor of 2.2. The VOP-file for the switchable array was constructed by concatenating the Q-matrices of the \say{up} and \say{down} states before VOP compression.
This procedure treats the two states of the array equivalently to two different voxel models. This ensures that all possible SAR distributions are safely represented by the jointly compressed VOP matrices.

\subsection{Numerical Pulse Optimization}
As a first proof of concept, the simulated \bplus maps were used for RF pulse optimization. A homogeneous flip angle distribution of 10\textdegree\ throughout the brain was chosen as the optimization target. The normalized root mean square error (nRMSE) of the obtained flip angle distribution was used as a quality metric. The optimization was performed based on the spatial domain method \cite{Grissom2006-le} using the variable exchange method \cite{Setsompop2008-ux} to solve the magnitude-least-squares problem with 20 iterations and no regularization.
Both single-channel operation in circularly polarized (CP) mode and 8-channel pTx operation were tested. For the switchable array, fast switching during the pulse was implemented by changing the coil profiles from “up” to “down” configuration after half the excitation time. In this case, magnitude and phase were optimized for each subpulse. 

For CP mode operation, one (reference array) or two (switchable array) global magnitudes and phases were calculated. For pTx operation, one or two RF shims were calculated instead. A 2-kT point excitation \cite{cloos2012kt} was also evaluated for both CP and pTx mode.
Note that, compared to the conventional kT points scenario in pTx mode, which requires suitable pTx hardware to apply variable RF shims for different subpulses, gradient blips between subpulses can also be used with CP mode excitation only. In this case, only the global magnitude and phase of each subpulse are variable, which can be executed using standard single-channel Tx hardware.

Since the focus of this work was to determine the effect of the switchable element, we tried to prevent any bias by the algorithm that determines the k-space locations of the kT-points. This was done by randomly generating 10,000 k-space positions in the range -14 m$^{-1}$ $\leq k_{x,y,z }\leq$ 14 m$^{-1}$ and selecting the one that gave the best performance for the given scenario as the first kT point. The second kT point was always positioned in the center of the k-space.

\subsection{Switchable Coaxial Dipole Array Prototype}
Based on the simulation results, an eight-channel 17 cm switchable array was constructed (\Cref{fig2} (A)). An optical trigger controlled by the pulse sequence was used for switching. The optical trigger was converted to a TTL-signal and then used to control a CMOS driver to switch the PIN diodes' states by applying $\pm$ 5 V to the control input of the switchable array. The switching signal was injected to the inputs of the switchable dipole elements using home-built bias-tees. The dipoles were constructed using semi-flexible non-magnetic RG-405 cable (Carlisle Interconnect). The dipole inputs and bias-tees outputs were connected using the same RG-405 cable with two floating ground cable traps \cite{bazooka_baloon} to prevent wave (common mode) propagation along the cable. The inputs of the bias-tees were connected to eight panel BNC connectors (Huber+Suhner, Switzerland) mounted on a polycarbonate flange plate. A photo of a single switchable unit consisting of two high-power PIN diodes (MA4P7446F-1091), lumped self-made 40 nH inductor (1.2 mm wire thickness) $L_{\text{end}}$, one DC-block capacitor $C_b$ and 2 RF-chokes is presented at \Cref{fig2} (C). Using two series-connected PIN diodes allows the protection of the switching unit from switching by the RF pulse itself.  The switchable array was tested in transceiver mode using a home built TxRx interface consisting of 8 LNAs (WantComm, USA) and 8 high-power TR-switches \cite{avdievich2017evaluation}. A detailed electric circuit of a single switchable antenna is presented in Figure \ref{fig2} (D). Switching time (~20\,µs) is defined by the PIN diode carrier lifetime.

\subsection{MRI Measurements and Pulse Design}

Experiments were performed on a Magnetom 9.4 T Plus whole-body MRI scanner (Siemens Healthineers, Erlangen, Germany). All experiments were performed with the approval of the local ethics committee and after obtaining written informed consent. The MR sequences were written in a modified version of Pulseq \cite{layton_pulseq_2017, freudensprung2023simple}, which allows pTx RF-shimming for each RF pulse. Single-channel \bplus and \bnull maps were acquired using a 3D presaturated turbo-flash sequence with weighted hybrid mapping \cite{bosch2023optimized}. \bplus and \bnull mapping was performed in a healthy male subject. The \bplus maps were acquired for both the \say{up} and the \say{down} configuration of the switchable array. The \bplus maps of the \say{up} and \say{down} configuration and their difference were compared to the equivalent maps from the simulation.

A homogeneous flip angle distribution of 10\textdegree\ throughout the brain was chosen as the optimization target. The nRMSE of the obtained flip angle distribution was optimized with 20 iterations of a Tikhonov-regularized minimization. Optimization was performed for the entire volume in the phantom and for the brain only in-vivo. Brain masks were generated from the proton-density weighted reference image of the presaturated turbo-flash sequence using SynthSeg \cite{billot_synthseg_2023}.
For excitation, 2-kT points pulses were designed. Fast switching during the pulse was implemented by changing the coil profiles from up to down configuration after half the excitation time. In this case, magnitudes and phases were optimized for each subpulse.

Similar to the simulation experiments, the k-space location of the first kT point was chosen by randomly generating 25,000 k-space positions in the range 28 m$^{-1} \leq k_{x,y,z} \leq$ 28 m$^{-1}$ and selecting the one that gave the best performance for the given scenario. The second kT point was again positioned in the center of the k-space.

\section{Results}
Numerically simulated \bplus maps in the central transversal plane of a homogeneous cylindrical phantom are presented in \Cref{fig1} (C) for the \say{down} and \say{up} cases. The ratio between these two \bplus profiles is presented at \Cref{fig1} (D), clearly illustrating the effect of skewed current densities on the resulting transmit sensitivity patterns.

\begin{figure}[htbp]
	\centering
	\includegraphics[width=0.9\textwidth]{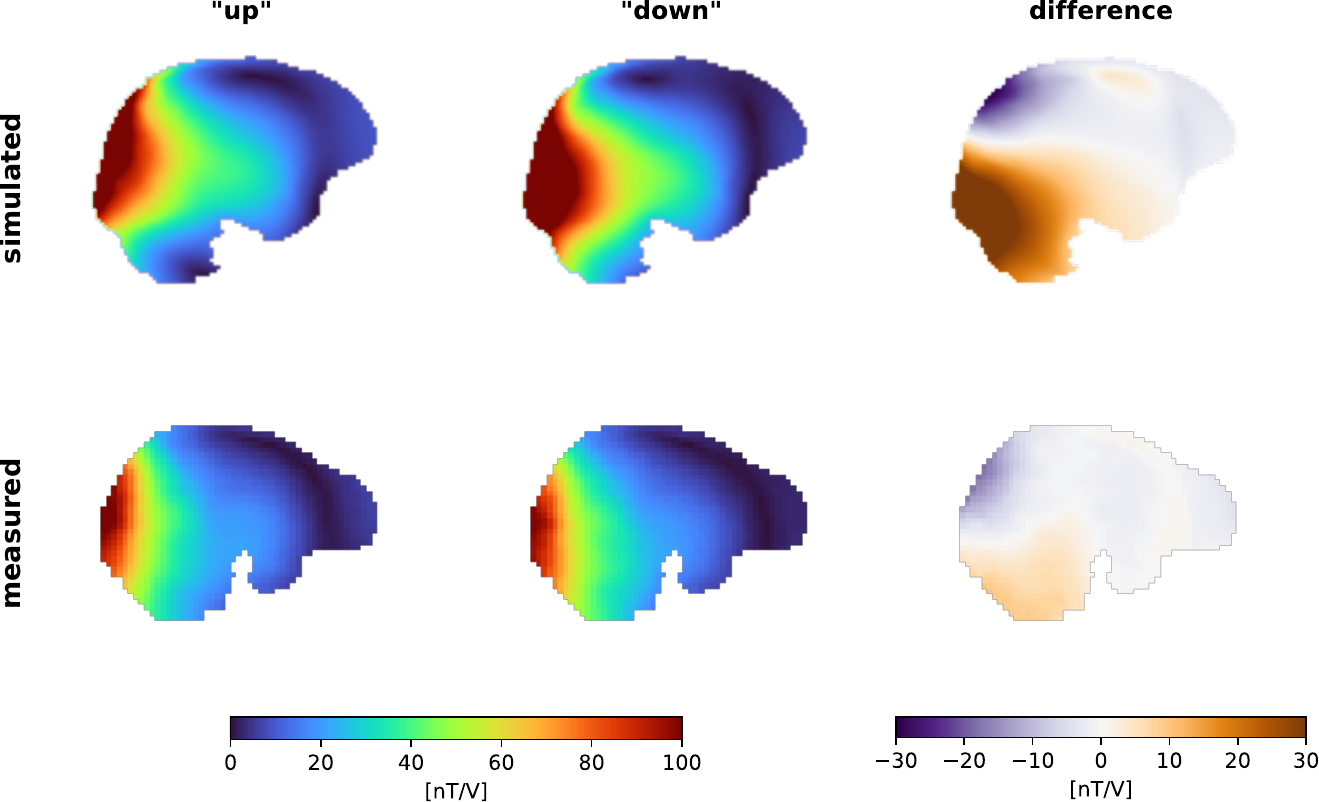}
	\caption{%
		Simulated and measured Tx sensitivities for both the \say{up} and \say{down} case.  While the switching has an effect, it is less pronounced than in the simulations.
	}%
	\label{fig3}
\end{figure}

As expected, the \bplus field distribution varied between the \say{up} and \say{down} states of the Tx elements also when integrated into an entire coil array. When switching from the \say{down} to the \say{up} configuration, the simulated \bplus field in the inferior regions got reduced by approximately 30\%, while the field in superior regions of the brain increased by a similar magnitude. In our experimental validation, this effect was also visible, but significantly lower than in the simulations. Here, the field in the inferior regions only reduced by ca. 10 \%. \Cref{fig3} displays the simulated and measured \bplus distribution of one element in both configurations, as well as the difference between the two configurations.

\begin{figure*}[tbp]
	\centering
	\includegraphics[width=0.9\textwidth]{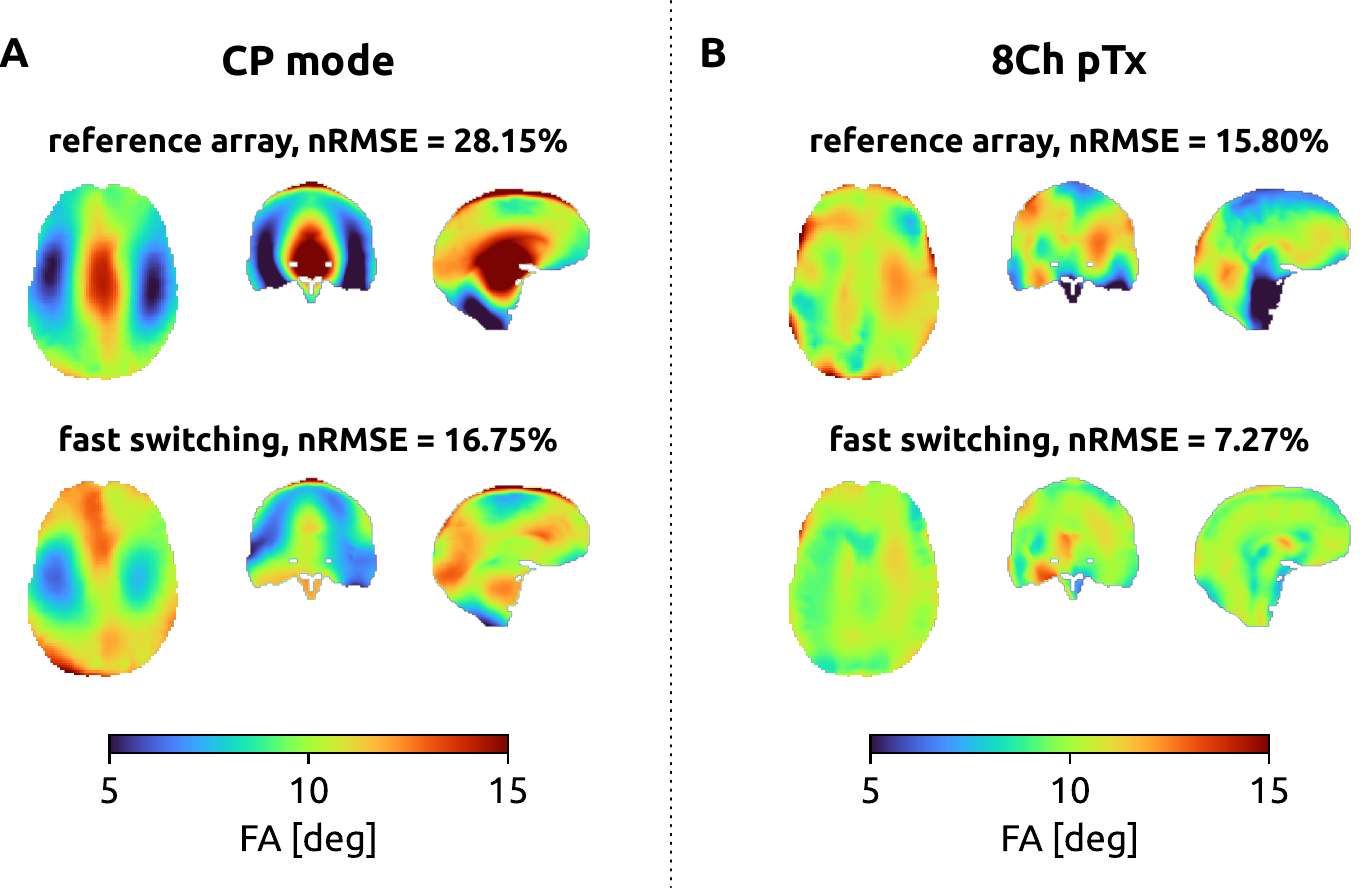}
	\caption{%
		Resulting flip angle distributions from simulations for 2 kT points pulses optimized on "Duke" in \textbf{(A)} CP mode operation, and \textbf{(B)} 8 channel pTx operation.
	}%
	\label{fig4}
\end{figure*}%

\subsection{Optimized RF Pulses Based on Simulated Coil}
When operating the coil in CP-mode, switching between the two coil sensitivities after half of a single excitation pulse improved the nRMSE from 32.47\% to 30.08\%. When using two kT points in CP mode, switching after half of every subpulse improved the nRMSE from 28.15\% to 16.75\% (\Cref{fig4} (A)).
When operating the coil in pTx mode, a single excitation pulse also benefited from fast sensitivity switching, as the nRMSE decreased from 29.47\% to 19.08\%. When designing a two kT pTx pulse, a nRMSE of 15.80\% (reference) and 7.27\% (switched) were achieved (\Cref{fig4} (B)).

\subsection{Optimized RF Pulses Based on Constructed Prototype Coil}
\begin{figure*}[tbp]
	\centering
\includegraphics[width=0.9\textwidth]{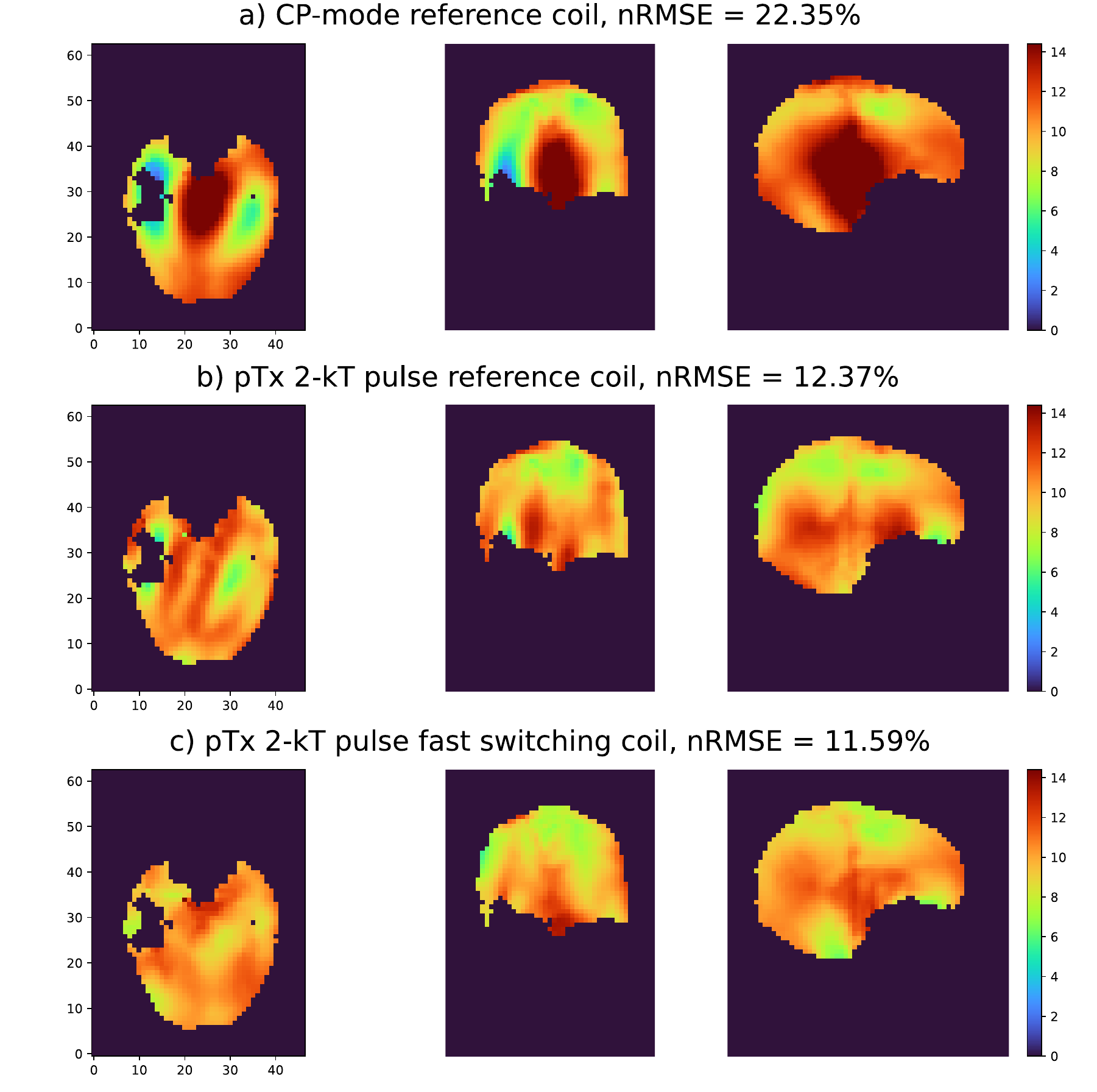}
\caption{%
		Resulting flip angle distributions from pulse optimization achieved on a subject. \textbf{(A)} Single-pulse CP mode configuration of the reference coil. \textbf{(B)} 2 kT points (pTx) pulse for the reference coil. \textbf{(C)} 2 kT points (pTx) with fast switching of transmit sensitivities.
	}%
	\label{fig5}
\end{figure*}

Based on the measured \bplus maps from the healthy volunteer, an nRMSE of 22.35\% was achieved for CP mode in the reference array (\Cref{fig5} (A)). The 2 kT points pTx pulse for the reference array achieved an nRMSE of 12.37\% (\Cref{fig5} (B)). Introducing sensitivity switching reduced the nRMSE to 11.59\% (\Cref{fig5} (C)).

\section{Discussion and Conclusion}
In simulations, introducing transmit elements with rapidly switchable \bplus sensitivities allows for a significant reduction in flip angle inhomogeneity both in CP and pTx mode operation (\Cref{fig4}).
However, compared to the simulation results, the constructed array prototype exhibits only a weak difference in the sensitivities of the \say{up} and \say{down} configurations, indicating that the effect of fast switching is well below expectations.
In contrast, previous work on similarly designed switchable Rx coaxial dipoles demonstrated a reasonable diversity between “up” and “down” configurations for both simulations and experimental results \cite{solomakha2024dynamic}.

A potential explanation for this disagreement is a suboptimal numerical model of the switchable unit. The entire switchable unit was simulated as an ideal single capacitive or inductive element. For the prototype, however, it was realized as a physical combination of different lumped elements soldered onto a PCB including relatively large PIN diodes required to switch high RF power. This addition of bulky PCB can lead to lower diversity between the \say{up} and \say{down} profiles.
First, the Tx dipole length was increased (from 13 cm to 17 cm) in comparison to Rx dipoles resulted in the smaller $L_{\text{end}}$ value. Including $L_{\text{end}}$ into a relatively large PCB required a further decrease of the inductance value making it comparable with the parasitic inductance of the entire PCB unit. In turn, in Rx dipoles, which didn’t require high RF power elements, the entire switchable circuit was substantially smaller. 
However, it is difficult to include the entire PCB in the numerical model, as this would lead to a very large simulation time. Next steps for improving upon that might be using different types of switchable elements (MEMS, MOSFETS), as well as other designs of switchable units that could reduce the size of the switchable unit itself.
Alternatively, the size of the dipoles can be reduced by combining two rows of sixteen shorter dipoles \cite{nikulin2023double}. This will also allow using smaller PIN diodes and increasing the $L_{\text{end}}$ value.

Still, in the course of the present work, we were able to demonstrate that the concept works not only in simulations, but in principle, it can also be realized experimentally for switching high RF power generated within the resonating Tx dipoles.
Similar to the observations made with reconfigurable Rx elements \cite{glang2022accelerated, nikulin2023reconfigurable, solomakha2024dynamic}, switching between different Tx sensitivity patterns of a single element can be seen as a way to effectively emulate a larger number of independent virtual Tx elements. Seeing the considerable cost and technical complexity of independent Tx channels, it becomes clear that the use of rapidly switchable \bplus sensitivities in transmit elements holds the promise to improve the degrees of freedom available for pulse design without costly modifications to the Tx chain. If the switching effect could be increased to that seen in the simulations, a large number of 8 channel pTx MRI systems could potentially be upgraded with a comparatively cheap hardware addition, which could drastically improve performance in areas that now prove difficult, such as combined whole-brain and C-spine imaging \cite{solomakha2025combined}.

\section{Acknowledgements}
Funding by the European Research Council (ERC Advanced Grant No 834940, SpreadMRI) and the DFG (HF-NeuroBOOST; DFG/ANR joint project; project number 530130666) is gratefully acknowledged.

\bibliography{sn-bibliography}

\end{document}